\documentclass{article}

\PassOptionsToPackage{numbers}{natbib}

\usepackage[preprint]{neurips_2023}

\usepackage[utf8]{inputenc} 
\usepackage[T1]{fontenc}    
\usepackage{hyperref}       
\usepackage{url}            
\usepackage{booktabs}       
\usepackage{amsfonts}       
\usepackage{nicefrac}       
\usepackage{microtype}      
\usepackage{xcolor}         
\usepackage{comment,amsmath,amssymb,subcaption,float,graphicx}
\hypersetup{colorlinks=true}
\usepackage[para,online,flushleft]{threeparttable}
\usepackage{multirow}

\newcommand\QD[1]{\textcolor{red}{[\textbf{QD: #1}]}}

\newcommand{\norm}[1]{\left\lVert#1\right\rVert}
\DeclareMathOperator{\E}{\mathbb{E}}

\title{RAND: Robustness Aware Norm Decay For Quantized Seq2seq Models}

\author{%
  David Qiu \qquad
   David Rim \qquad
   Shaojin Ding \qquad
   Oleg Rybakov \qquad
   Yanzhang He \\
  Google, LLC\\
  \{qdavid, davidrim, shaojinding, rybakov, yanzhanghe\}@google.com
}

\begin{document}

\maketitle

\begin{abstract}
With the rapid increase in the size of neural networks, model compression has become an important area of research. Quantization is an effective technique at decreasing the model size, memory access, and compute load of large models. Despite recent advances in quantization aware training (QAT) technique, most papers present evaluations that are focused on computer vision tasks, which have different training dynamics compared to sequence tasks. In this paper, we first benchmark the impact of popular techniques such as straight through estimator, pseudo-quantization noise, learnable scale parameter, clipping, etc. on 4-bit seq2seq models across a suite of speech recognition datasets ranging from 1,000 hours to 1 million hours, as well as one machine translation dataset to illustrate its applicability outside of speech.

Through the experiments, we report that noise based QAT suffers when there is insufficient regularization signal flowing back to the quantization scale. We propose low complexity changes to the QAT process to improve model accuracy (outperforming popular learnable scale and clipping methods). With the improved accuracy, it opens up the possibility to exploit some of the other benefits of noise based QAT: 1) training a single model that performs well in mixed precision mode and 2) improved generalization on long form speech recognition.

\end{abstract}

\section{Introduction}




Sequence-to-sequence (seq2seq) model is an influential class of neural architecture in various research fields and real-world AI systems. Due to the emerging demands from user-interactive devices and services (e.g., search by voice, voice assistant, etc.), end-to-end (E2E) automatic speech recognition (ASR)~\cite{wang2019overview, hannun2014deep, graves2012sequence, chorowski2015attention, dong2018speech, li2020comparison,he2019streaming,CC18,KimHoriWatanabe17,JinyuLi2019,Zeyer2020} has been widely investigated and has seen dramatic quality improvements in the past few years. With the emergence of large models and the limited budgets on hardware, reducing the inference and training cost through model compression is a core problem for these devices and services. 

Quantization reduces the number of bits required to represent weight tensors and activations, and is an effective technique at decreasing the model size, memory access, and compute load of large models. A standard approach of applying model quantization is through post-training quantization (PTQ) with int8~\cite{POSTQUANT}. At int4 or lower precision, PTQ typically shows a large gap in performance compared to float inference~\cite{abdolrashidi2021pareto}. Quantization aware training (QAT)~\cite{nguyen2020quantization, prasad2020quantization, ding20224} amends the training process with noisy operations that models the quantization errors encountered during inference, and is usually needed to close the gap.

Existing QAT techniques are generally implemented using real quantization noise or pseudo-quantization noise (PQN). Injecting real quantization noise rounds the weights to integer precision during training, and requires a straight through estimator (STE)~\cite{bengio2013estimating} to bypass the round function during backpropagation~\cite{courbariaux2016binarized}. STE QAT has been observed to cause weights to not converge and oscillate around quantization decision boundaries~\cite{defossezdifferentiable}. On the other hand, PQN QAT exposes weights to a large amount of noise even when the weight would incur zero quantization rounding error. One way of reducing the noise variance needed to achieve quantization robustness is to introduce extra learnable scale parameters that can be regularized in an E2E fashion~\cite{esserlearned,baskin2021nice,jain2020trained,park2022nipq}. Since the learned scale cannot be guaranteed to cover the entire range of weight values, clipping must be introduced. We denote these methods as learnable scale and clip (LSC).

Instead, we advocate for using a per-channel $L_p$ norm of the weight tensor as the scale, such that it removes the need for clipping. This makes PQN QAT free of any STE. As most of the edge devices are memory-bounded, we mainly focus on weight quantization in this work, though the techniques presented can be extended to activation quantization as well. Overall, this paper makes the following contributions:
\begin{enumerate}
\item Share benchmark results for QAT on multiple large scale speech recognition and one machine translation seq2seq models and datasets, which expands the mostly computer vision and language modeling focused QAT literature.
  
\item Show that PQN QAT is more sensitive to scale (i.e. the size of the hyper-rectangle that we are optimizing the loss over), such that the gap with and without scale regularization is bigger compared to that of STE QAT.
  
\item Propose robustness aware norm decay (RAND), which directly uses the weight matrix's per-channel $L_p$ norms as the quantization scales and decay those norms in an E2E QAT procedure. The performance of RAND improves as we move from per tensor scale to per channel scale, and from single-domain to multi-domain models and tasks, beating LSC.
  
\item Demonstrate how RAND improves the performance of post-training selection of layer precision to enable mixed precision inference without additional training time complexities.
  
\item Show that RAND has generalization benefits on long form caption tasks beyond traditional regularization techniques such as variational noise (VN)~\cite{graves2011practical}.
\end{enumerate}

\section{Methods}
\subsection{Preliminaries}
A simple fully connected matrix multiplication can be written as $Y = WX$, where $Y \in \mathbb{R}^M$, $X \in \mathbb{R}^N$, and $W \in \mathbb{R}^{M \times N}$. Quantizing the entire weight matrix with a single scale parameter $s \in \mathbb{R}$ involves: 1) dividing by $s$ to convert from float to int range, 2) rounding to the nearest integer, 3) clipping to the integer precision lower bound $l$ and upper bound $u$. For simplicity, we focus on symmetric uniform quantization ($l=-7$ and $u=7$ for 4-bit quantization). Multiplication with a quantized weight matrix can be modeled as
\begin{equation}
    \label{Eq: per tensor}
    Y = s  \cdot \left[ \text{clip} \left( \text{round} \left(\frac{W}{s}\right) , l, u \right) X \right].
\end{equation}

To combat the effect of outliers, every output channel can have its dedicated scale. Then, matrix multiplication can be modeled as
\begin{equation}
\label{Eq: per channel}
    Y_i = s_i  \cdot \left[ \text{clip} \left( \text{round} \left(\frac{W_{i}}{s_i} \right) , l, u \right) X \right], 1 \leq i \leq M,
\end{equation}
where $s_i \in \mathbb{R}$ and denotes the $i$-th channel's scale, and $W_i$ denotes the $i$-th row of $W$. Common QAT methods optimizes the neural network by using~\eqref{Eq: per tensor} or~\eqref{Eq: per channel} in the forward propagation, and a straight through estimator (STE)~\cite{bengio2013estimating} with carefully designed gradients to bypass the clip and round functions in the backpropagation. The quantized weights can be materialized during training or created during an additional post-training quantization step.

\subsection{Robustness Aware Norm Decay For Quantization Aware Training}
It is well known from signal processing~\cite{widrow1996statistical} and neural compression~\cite{agustsson2020universally} that the uniform distribution does a good job in modeling quantization noise. Since the maximum noise introduced by the round function is $\frac{1}{2}$, which is then amplified by $s_i$, training with uniform noise drawn from $\text{Unif}\left[-\frac{s_i}{2}, \frac{s_i}{2} \right]$
added to the weights is an effective way to emulate the QAT process. This removes the need for a STE to bypass the round function, which has zero gradient almost everywhere. Training with this uniform pseudo-quantization noise (PQN) changes~\eqref{Eq: per channel} into
\begin{equation}
\label{Eq: per channel noise}
    Y_i = (W_i+ s_i Z_i) X, 1 \leq i \leq M,
\end{equation}
where $Z_i \in \mathbb{R}^M$ and every entry is drawn from from $\text{Unif}\left[-\frac{1}{2}, \frac{1}{2} \right]$. An additional benefit of PQN-QAT is the interpretability of the learning problem, which, when focusing on $W_i$, becomes
\begin{equation}
\label{Eq: optimization}
    \min_{W_i} \E_{Z_i} L(f_{W_i+s_i Z_i}),
\end{equation}
where $L(\cdot)$ denotes the training loss over the entire dataset. Essentially, \eqref{Eq: optimization} finds the optimal $W_i$ where the average loss over a hyperrectangle centered at $W_i$ with lengths of $s_i$ in all dimensions, is minimized (see Figure~\ref{Fig: landscape}). This is in agreement with the desirable convergence to flat minima from generalization literature~\cite{li2018visualizing,du2022sharpness}, and we explore the secondary generalization benefits offered by our proposed method in Section~\ref{Section: prod data}.

However, the noise introduced by the clip function in~\eqref{Eq: per channel} is not accurately modeled by uniform noise. Re-introducing clip into~\eqref{Eq: per channel noise} again requires a STE with properly designed gradients. For \textit{robustness aware norm decay (RAND)}, we advocate for eliminating the extra trainable scale parameter $s_i$ and directly computing it as a function of the network weights. In particular, set $s_i \triangleq c \norm{W_i}_p$ in~\eqref{Eq: per channel noise}, where $c$ is a constant chosen for the entire neural network, and $\norm{\cdot}_p$ is the vector $L_p$ norm.
\begin{equation}
\label{Eq: rand}
    Y_i = (W_i+ c\norm{W_i}_p Z_i) X, 1 \leq i \leq M
\end{equation}
is the generalized equation for matrix multiplication under RAND.

\subsection{Training Modes for Robustness Aware Norm Decay}
\begin{figure}
    \centering
    \begin{minipage}{0.47\textwidth}
        \centering
        \includegraphics[width=1\textwidth]{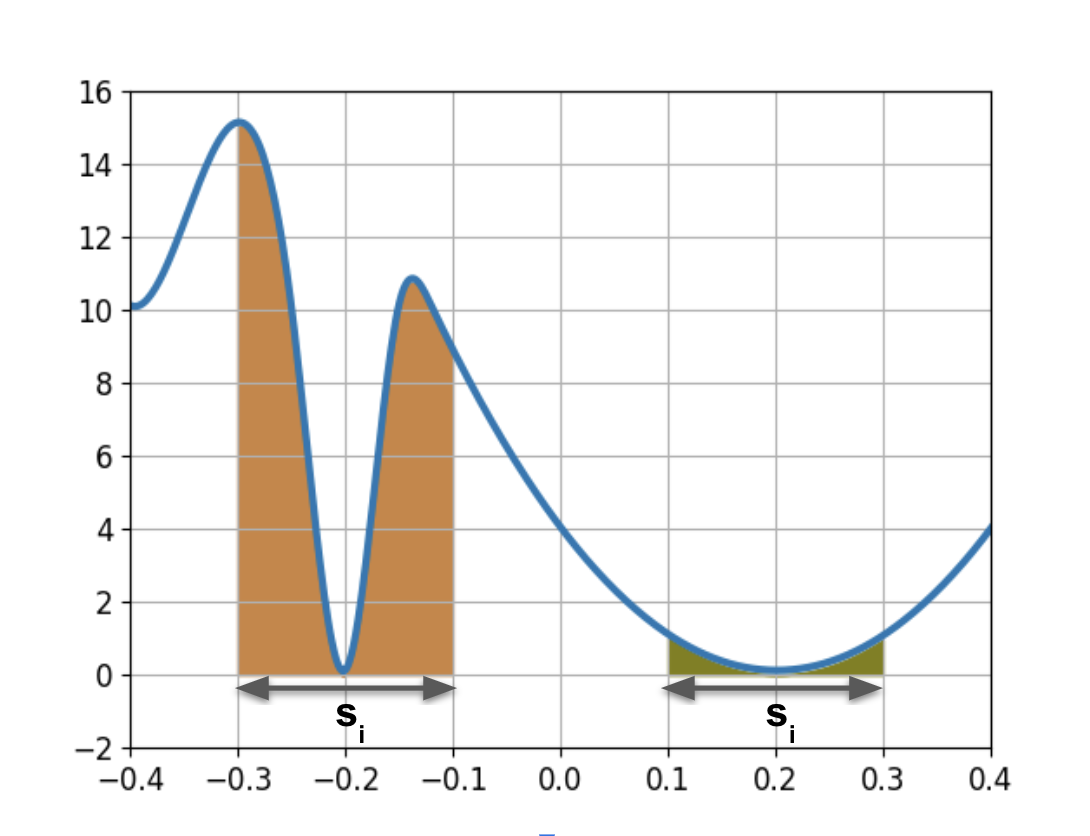}
        \caption{1-D example of two possible training losses (shaded areas). RAND training with noise prefers the flat minimum on the right. Norm decay attenuates $s_i$ in an E2E fashion.}
        \label{Fig: landscape}
    \end{minipage}\hfill
    \begin{minipage}{0.47\textwidth}
        \centering
        \includegraphics[width=1\textwidth]{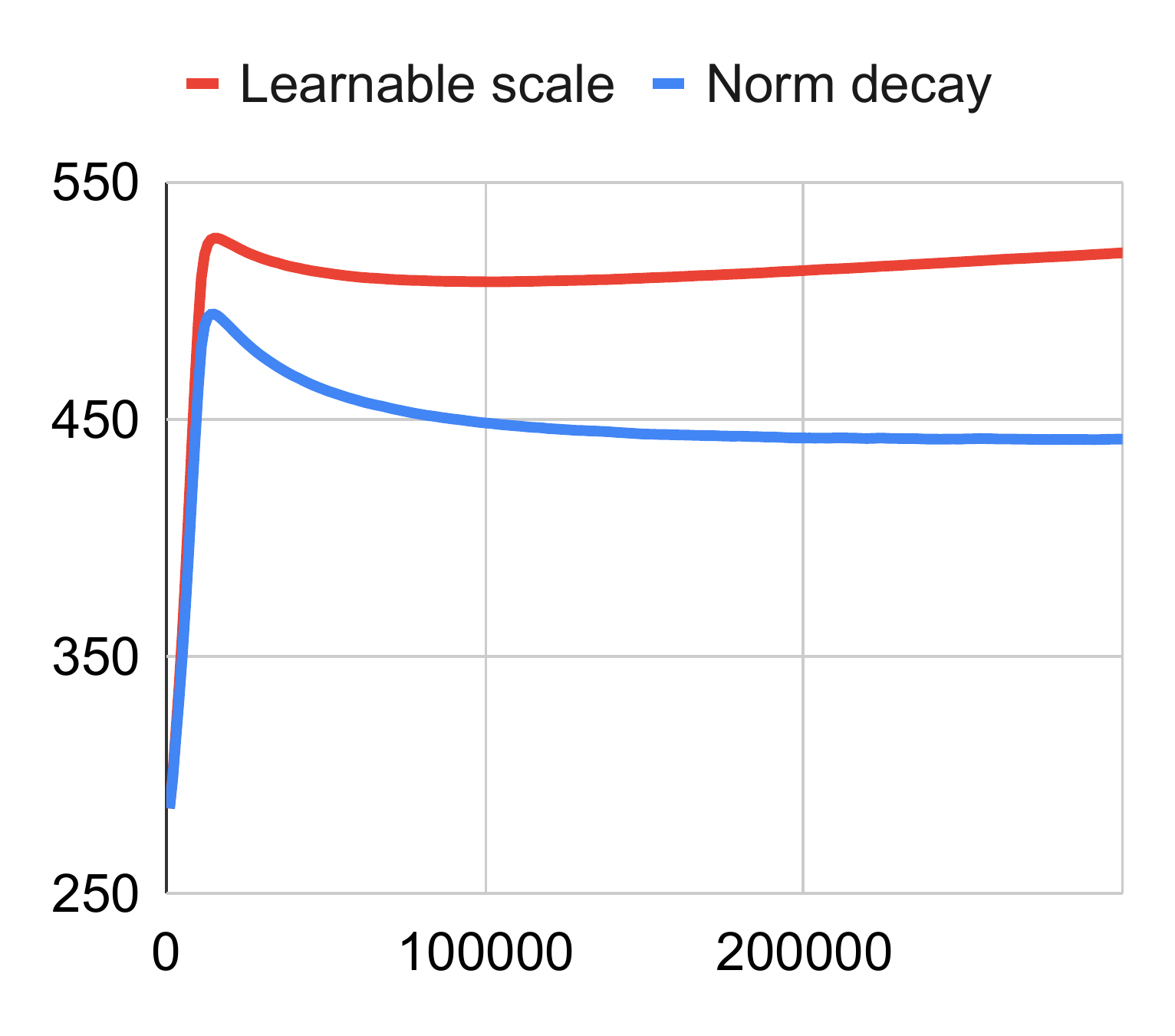}
        \caption{The sum of square of all network weights as a function of training iterations, for the ConformerS architecture on LibriSpeech.}
        \label{Fig: weights}
    \end{minipage}
\end{figure}

For~\eqref{Eq: rand}, we highlight three representative settings for $p$ and $c$:
\begin{align}
\label{Eq: inf norm}
\textbf{QAT mode: } &p=\infty, c = \frac{1}{2^{\text{bit}-1}-1} \\ 
\label{Eq: large norm}
\textbf{Mixed scale QAT mode: } &2 < p \leq \infty, 0 \leq c \leq  \frac{1}{2^{\text{bit}-1}-1} \\
\label{Eq: 2 norm}
\textbf{Generalization mode: } &p = 2, c \geq 0.
\end{align}
\textbf{QAT mode:} In this paper, we focus on QAT mode, which sets $s_i \triangleq \frac{\max_j |W_{i,j} |}{2^{\text{bit}-1}-1}$ to fully cover the range of $W_i$ such that the clip function becomes a no-op. This converts~\eqref{Eq: rand} into
\begin{equation}
\label{Eq: matmul inf norm}
    Y_i = \left(W_i+ \frac{\text{maybe\_stop\_gradient}( \max_j |W_{i,j} |)}{2^{\text{bit}-1}-1} Z_i \right) X, 1 \leq i \leq M.
\end{equation}
When maybe\_stop\_gradient is set to true, \eqref{Eq: matmul inf norm} only uses $\max_j |W_{i,j} |$ to determine the maximum amount of quantization noise that the network needs to be exposed to. When maybe\_stop\_gradient is set to false, the loss gradient with respect to the noisy perturbation directly informs how much $\max_j |W_{i,j} |$ (i.e., the biggest outlier) should be decayed. We call this \textit{robustness aware norm decay} because the uniform weight noise introduces perturbation robustness into the learned set of weights (i.e., prefer the flatter minimum in Figure~\ref{Fig: landscape}), and simultaneously the per channel $L_p$ norm of $W$ is being decayed depending on how sensitive that output channel is to weight perturbations (i.e., decreases the width of the shaded region that the loss is averaged over in Figure~\ref{Fig: landscape}).

\textbf{Mixed scale QAT mode:} QAT mode performs well when there is per channel (or per sub-channel) scale. However, in certain situations (e.g., with per tensor scale, or a large number of weights per channel), only one weight entry out of the entire weight tensor receives gradients from the noisy perturbation. In modern, large neural networks, the weight matrices may have $10^5$ or more entries. When one outlier gets attenuated, it is highly probable that there is another one with similar magnitude, and norm decay would struggle to fully regularize all large weight values. We report experimental results to illustrate this effect in Sections~\ref{Section: per tensor scale} and~\ref{Section: per channel scale}. As an alternative, replacing the infinity norm in QAT mode with \eqref{Eq: large norm} softens the dependency on a single max value and allows the gradients to attenuate multiple outliers per training iteration. We experiment with this setting in the Appendix.

\textbf{Generalization mode:} \eqref{Eq: 2 norm} and \eqref{Eq: inf norm} can be viewed as extensions of training with noisy weights (e.g., variational noise~\cite{graves2011practical}). Intuitively, output channels corresponding to large weight values contain features with higher importance and are more sensitive to small input perturbations. As training proceeds, the relative effect of a constant weight noise variance on output channels with increasing weights will decrease. Scaling the noise by the norm of the corresponding weight vector allows the noise variance to update with the weights.

\section{Related Works}


\subsection{Neural Network Quantization and Quantization Aware Training}

To decrease the latency and model size without compromising recognition quality, network quantization has been widely explored on ASR models~\cite{han2015deep, alvarez2016efficient, he2019streaming, prasad2020quantization, nguyen2020quantization, kim2021integer, ding20224}. In general, most of the existing research focus on weight and/or activation quantization. Weight quantization can save memory footprints on devices, while activation quantization can further improve computational efficiency by using integer multiplication. Among model quantization methods, post-training quantization (PTQ) with int8 is popular and easy to use for inference on edge devices~\cite{POSTQUANT}. It is successfully applied in multiple applications~\cite{he2019streaming, sainath2020streaming}. Advanced PTQ methods (e.g., optimal brain compression) seek to exploit the correlations in the feature activations to determine the optimal quantization rounding policy~\cite{nagel2020up, hubara2020improving, frantaroptimal}. These works are complementary to QAT and our work.

For lower-bit quantization, QAT is usually needed to mitigate the loss of precision, as shown in~\cite{nguyen2020quantization, prasad2020quantization, ding20224}. These popular QAT techniques expose the network to quantization errors, and use STE to bypass the quantization round function, which has zero gradient almost everywhere. Our work seeks to improve the QAT procedure to be free of STE and can be combined with recent advances in PTQ for additional improvements.

\subsection{Neural Network Training with Noisy Perturbations}
Neural networks can be trained with either noisy perturbations on the weights or features for the goal of generalization, quantization robustness, or adversarial robustness. For example, variational noise~\cite{graves2011practical} reformulates neural network training as Bayesian inference, and gives a Bayesian interpretation to training with Gaussian noise added to every weight. It has been widely adopted in the ASR community for improving generalization~\cite{gulati2020conformer,li2021confidence}. Our work and other noise based quantization works~\cite{baskin2021nice,defossezdifferentiable,park2022nipq} can be viewed as modifying the VN distribution to be specific to the uniform quantiztion noise introduced at each weight tensor, though this loosens the Bayesian connection.

One benefit of PQN-QAT over STE-QAT is that it prevents weight oscillation near the quantization decision boundary~\cite{nagel2022overcoming}. \cite{fan2020training} documents this drawback of STE and proposed to stochastically select subsets of weights to remain at full precision during every training iteration. However, the quantized weights are still trained with STE. \cite{baskin2021nice} (and~\cite{baskin2021uniq} for non-uniform quantization) proposes to use uniform noise in the QAT process, but the paper advocates to slowly progress to STE-QAT as the training proceeds. ~\cite{defossezdifferentiable} uses uniform noise throughout the training, and is comparable to our work. However, the paper's exposition on the training procedure is done through an example where weights are normalized to the range of $[0, 1]$, which does not make it clear whether the weight tensor's norm can be part of the optimization to control for outliers. Additionally,~\cite{defossezdifferentiable} focuses on a single quantization scale per weight tensor, which we show is sub-optimal for noise based QAT (see Section~\ref{Section: per tensor scale}). To our knowledge, our work reports new findings to the field by fully analyzing the effect of PQN-QAT, per-channel scale, and end-to-end norm decay on seq2seq tasks.

Similar to QAT's goal for \textit{weights}, adversarial robustness methods aim to train a network that is robust to \textit{input features} that are subject to a worst case perturbation within a pre-defined $L_\infty$ ball. Classic methods for generating these adversarial examples include the fast gradient sign method~\cite{goodfellow2014explaining} and projected gradient descent~\cite{madrytowards}. Although they are outside the scope of this paper, we remark that adversarial robustness methods may be adapted to help QAT.

\subsection{Quantization Aware Training with Learnable Scale}
Instead of using a $L_p$ norm as the quantization scale, many works (\cite{esserlearned,baskin2021nice,jain2020trained,park2022nipq}) advocate for creating extra learnable parameters as either the scale or clipping bounds. The gradients for weight entries that are clipped requires a STE, which can cause those weights to increase in magnitude even though they are already outside of the clipping bounds. Figure~\ref{Fig: weights} plots the $L_2$ norm of all weights after training with learnable scale and our propose method to compare the difference in behaviors. Additionally, the highly discontinuous nature of the gradients flowing to the learnable scale (see equation 3 of~\cite{esserlearned}) can introduce training instabilities and causes our multi-domain experiments to diverge (see Section~\ref{Section: stew}). For ASR, we use empirical results to argue that RAND outperforms learnable scale and clipping (LSC) methods.

\section{Experimental Setup}

\subsection{Datasets}

We conduct experiments on three ASR datasets of different scale and one machine translation dataset (MT) to systematically evaluate the proposed approach: 1) LibriSpeech~\cite{panayotov2015librispeech} (960 hour single-domain); 2) SpeechStew~\cite{chan2021speechstew} (5,000 hour multi-domain); 3) An in-house dataset (1 million hour multi-domain); 4) WMT En-Fr (see Appendix 1 for the full setup and results). We use word error rate (WER) as the evaluation metric for all the ASR experiments.

\textbf{LibriSpeech} training set contains 960 hours of speech (460 hours “clean” speech and 500 hours “noisy” speech). The test set also consists of “clean” and “noisy” versions. 

\textbf{SpeechStew} has a mix of publicly available datasets from different domains, including AMI~\cite{kraaij2005ami}, Broadcast News, Common Voice~\cite{ardila2019common}, LibriSpeech~\cite{panayotov2015librispeech}, Switchboard/Fisher, TED-LIUM~\cite{rousseau2012ted, hernandez2018ted}, and Wall Street Journal. Following the protocol in~\cite{chan2021speechstew}, we train the models using the mixed training set, and evaluate it on each individual test set.

Our \textbf{in-house training set} consists of over 1 million hours of English speech utterances from multiple domains, including voice assistant, voice typing, video captioning, etc. Most utterances in the training set are transcribed with a teacher model, 
while a small number of utterances are anonymized and hand-transcribed. We use three test sets from the voice assistant, voice typing, video captioning domains for evaluation. All of the test sets contain anonymized and hand-transcribed utterances.

\subsection{Backbone ASR Architecture and Implementation Details}

\begin{table}
  \caption{The number of parameters and model sizes after encoder quantization for each backbone archiecture. For ConformerL and ConformerS, the decoder remains in float. For Cascaded Conformers, the decoder is post-training quantized to int8.}
  \label{Table: backbone_size}
  \centering
      \begin{tabular}{llll}
        \toprule
        Backbone & \#Params & Size with int8 encoder  &  Size with int4 encoder \\
        \midrule
        ConformerL & 118 Million & 138 MB  & 81 MB    \\
        ConformerS & 10 Million & 16 MB  & 12 MB    \\
        Cascaded Conformers & 122 Million & 123 MB  & 65 MB    \\
        \bottomrule
      \end{tabular}
\end{table}

We considered two ASR backbones in our experiments. For the experiments on LibriSpeech and SpeechStew, we implemented \textbf{ConformerL} and \textbf{ComformerS} as proposed in~\cite{gulati2020conformer} to obtain better reproducibility and fair comparisons with prior studies. For the experiments on the in-house data, we implemented the a large-medium \textbf{Cascaded Conformers} as proposed in~\cite{ding2022unified} for optimal WER and latency under real world settings. The total number of parameters and model sizes after quantizations are shown in Table~\ref{Table: backbone_size}.

\textbf{ConformerL} and \textbf{ConformerS} have a frontend of 80-dimensional log Mel-filterbank energies, extracted from 25ms window and 10ms shift. The two variants have 17,  16 conformer layers, with 512, 144 dimensions, respectively. In each conformer layer, both have 32-dimensional kernel in depthwise convolutions, while ConformerL has 8 attention heads in self-attention and ConformerS has 4. The LSTM layer in the RNN-T prediction network has 640 units.

The \textbf{Cascaded Conformers} have a frontend of 128-dimensional log Mel-filterbank enegies. In addition, the 4 contiguous frames are stacked, which is then sub-sampled by a factor of 3. The model is comprised of a causal encoder and a non-causal encoder, along with separate RNN-T decoders for each encoder. The causal encoder has 9 conformer layers, where there is no self-attention in the first 3 layers. Each layer has 23-frame left context per layer and no right context to strictly prevent the model from using future inputs. The non-causal encoder has 6 conformer layers, with a cumulative 900ms of right context. All of the self-attention layers have 8 heads, and all layers use causal convolution with a kernel size of 15. Each separate RNN-T decoder consists of an 320-dimensional embedding prediction network and a 384-dimensional fully-connected joint network. 

\subsection{Training and Evaluation Details}

For all models, we experiment with RAND and four other competing methods that we reproduce to the best of our knowledge (referenced in Table~\ref{Table: libri per channel}). We also adapt the norm decay part of RAND to STE-QAT and report results for it. We quantize only the encoders since the they represent the overwhelming majority of parameter for ASR models. Additionally, we do not quantize convolutional layers due to their lower parameter count relative to fully connected layers.

All models are trained on Tensor Processing Unit (TPU) v3-128~\cite{TPU} with the Adam optimizer~\cite{kingma2014adam}. ConformerS and ConformerL use a base learning rate of $5.0$, multiplied by the transformer learning rate schedule~\cite{vaswani2017attention} with warmup steps set to $10000$, and a batch size of 2048. The Cascaded Conformer use a base learning rate of $7.5$, with warmup steps set to $32000$, and a batch size of 4048.

LibriSpeech ConformerS models are trained to 300,000 steps and evaluated at the checkpoint with the best dev set WER. SpeechStew ConformerL and the Cascaded Conformer do not overfit on their multi-domain training data, and we evaluate at exactly 200,000 steps and 700,000 steps, respectively. All evaluations use an exponential moving average version of the network weights, computed with a decay factor of $0.9999$.


\section{Results}

\subsection{Preliminary Evaluation: LibriSpeech with Per Tensor Scale}
\label{Section: per tensor scale}
To quantify the effect of the norm decay method within RAND, we first experiment with the different QAT techniques on a ConformerS model with per tensor scale. When each weight tensor only has a single scalar as the scale, any outlier affects the dynamic range of the entire tensor. Although this setup is sub-optimal for 4-bit and lower quantization, it accentuates the differences between different outlier control methods and is useful for analysis.

Table~\ref{Table: libri per tensor} shows the effectiveness of norm decay on both STE-QAT and PQN-QAT. The Test-other WER are reduced from 8.41 and 9.74 to 6.43 and 6.50, respectively. The larger improvement suggests that norm decay is especially necessary for PQN-QAT. Training with weight noise can be interpreted as sacrificing model capacity to induce model robustness to perturbations in any direction. That is a stronger robustness guarantee than STE-QAT, which only aims to be robust to the direction of the inference time quantization algorithm. For example, PQN-QAT can cause weights near zero to change signs, which is impossible under STE-QAT. Thus, it is expected that PQN-QAT needs norm decay to lower the amount of perturbation the model needs to be robust to.

\begin{table}
  \caption{ConformerS experiments for 4-bit symmetric weight quantization with per tensor scale, on LibriSpeech.}
  \label{Table: libri per tensor}
  \centering
  \begin{tabular}{lllll}
    \toprule
    Eval precision & Outlier method    & QAT method     & Test-clean & Test-other \\
    \midrule
    Float & None  & None & 2.51 & 6.15    \\
     Int4     & None & 4-bit STE & 3.53 & 8.41      \\
     Int4   & None & 4-bit PQN & 3.94 & 9.74  \\
     Int4     & LSC & 4-bit STE & 2.71 & 6.66      \\
     Int4   & LSC & 4-bit PQN & 2.73 & 6.30  \\
     Int4     & Norm & 4-bit STE & 2.74 & 6.43      \\
     Int4   & Norm & 4-bit PQN & 2.71 & 6.50  \\
    \bottomrule
  \end{tabular}
\end{table}

\subsection{Single-domain Evaluation: LibriSpeech with Per Channel Scale}
\label{Section: per channel scale}
When we compare the WER between LSC and norm decay in Table~\ref{Table: libri per tensor}, there is no conclusive trend. Under per tensor scale, one major problem that norm decay faces is that only one weight value can serve as the scale during each training iteration. In modern, large neural networks, the weight matrices may have $10^5$ or more entries. When one outlier gets attenuated, it is highly likely that there is another one with similar magnitude, and norm decay would not have enough iterations to fully regularize all weight values.

In this subsection, we report the LibriSpeech WER in a more realistic setting for 4-bit: when the model is quantized with per channel scale. Now, in every iteration, every channel's maximum value is regularized by QAT. Table~\ref{Table: libri per channel} shows that with per-channel scale, norm decay with STE-QAT or PQN-QAT show lower WER than their LSC counterparts, with PQN-QAT (RAND) outperforming STE-QAT.

\begin{table}
  \caption{ConformerS experiments for 4-bit symmetric weight quantization with per channel scale, on LibriSpeech, reported as WER $\pm$ standard deviation.}
  \label{Table: libri per channel}
  \centering
  \begin{tabular}{llllll}
    \toprule
    Reference & Eval precision & Outlier method    & QAT method     & Test-clean & Test-other \\
    \midrule
    N/A & Float & None  & None & 2.51 & 6.15     \\
    \cite{ding20224} & Int4     & None & 4-bit STE & 2.83$\pm$0.02 & 6.58$\pm$0.06      \\
    \cite{defossezdifferentiable} & Int4   & None & 4-bit PQN & 2.82$\pm$0.03 & 6.77$\pm$0.08  \\
    \cite{esserlearned} & Int4     & LSC & 4-bit STE & 2.78$\pm$0.03 & 6.53$\pm$0.10      \\
    \cite{park2022nipq} & Int4   & LSC & 4-bit PQN & 2.63$\pm$0.01 & 6.46$\pm$0.07  \\
    RAND & Int4     & Norm & 4-bit STE & 2.73$\pm$0.04 & 6.34$\pm$0.03      \\
    RAND & Int4   & Norm & 4-bit PQN & 2.64$\pm$0.02 & 6.30$\pm$0.04  \\
    \bottomrule
  \end{tabular}
\end{table}

\subsection{Multi-domain Evaluation: SpeechStew Dataset}
\label{Section: stew}
LibriSpeech is generally viewed as an easier dataset where models exhibit overparameterization properties. The redundancy benefits offered by overparameterization generally helps the network be robust to random weight perturbations. To extend the empirical analysis, we report WER results for ConformerL evaluating on the multi-domain dataset SpeechStew, which is a more diverse and difficult task.

In Table~\ref{Table: stew}, we see the same trend where, as an outlier control method, norm decay mostly outperforms LSC across the range of tasks. Interestingly, for these larger models, STE-QAT with LSC does not converge. We hypothesize that the highly discontinuous gradients that arises due to the clipping function hurts the convergence of the learnable scale parameters (see equation 3 of~\cite{esserlearned}). By contrast, norm decay training remains stable for larger models. Under RAND's norm decay, PQN-QAT outperforms STE-QAT by a small margin in terms of the average WER over all datasets, but we report the additional generalization and multi-precision benefits of RAND with PQN-QAT in the Section~\ref{Section: prod data}.

\begin{table}
  \caption{Multi-domain ConformerL experiments for 4-bit symmetric weight quantization with per channel scale, on SpeechStew dataset. CV: Common Voice; SB: Switchboard; LS: LibriSpeech; Ted: TED-LIUM; WSJ: Wall Street Journal.}
  \label{Table: stew}
  \centering
  \begin{threeparttable}
  \begin{tabular}{lll|cccccccc}
    \toprule
    Eval & Outlier    & QAT & \multicolumn{2}{c}{AMI} & \multirow{2}{*}{CV}  & \multicolumn{2}{c}{LS-Test} & \multirow{2}{*}{SB} & \multirow{2}{*}{Ted} & \multirow{2}{*}{WSJ}         \\
    \cmidrule(r){4-5} \cmidrule(r){7-8}
    Prec &  Method   &   Method   & IHM & SDM1 &  & clean & other \\
    \midrule
    
    Float & None  & None & 9.19 & 23.53 & 9.89 & 2.03 & 4.32 & 8.63 & 3.98 & 1.38  \\
    Int4     & LSC & 4-bit STE\tnote{*} & -- & -- & --  & -- & -- & -- & -- & --   \\
    Int4   & LSC & 4-bit PQN & 9.33 & 24.31 & 10.48 & 2.17 & 4.55 & 8.87 & 4.44 & 1.58 \\
    Int4     & Norm & 4-bit STE & 9.22 & 24.33 & 10.18 & 2.07 & 4.38 & 8.45 & 4.45 & 1.55    \\
    Int4   & Norm & 4-bit PQN & 9.39 & 24.23 & 10.17 & 2.07 & 4.48 & 8.58 & 4.11 & 1.44 \\
    \bottomrule
  \end{tabular}
    \begin{tablenotes}
        \item[*] Model failed to converge.
    \end{tablenotes}
  \end{threeparttable}
\end{table}


\subsection{Large Scale Evaluation: In-house Dataset}
\label{Section: prod data}
In a real world production setting, training different models per domain and per platform causes scaling and maintainability challenges. Ideally, the same ASR model can be used for short form (e.g., voice assistant), medium form (e.g., voice typing), and long form (e.g., video captioning), as well as multiple hardware platforms with differing support for int4, int8, and float quantization schemes. We report WER on a real world, large scale in-house dataset. As Section~\ref{Section: stew} established norm decay as the stronger outlier control method over LSC, we now focus on the norm decay method and compare STE and PQN-QAT under this multi-domain, multi-precision setting.

\textbf{Generalization:}
Long form data tends to have higher variance in terms of background noise, pause / silence / segmentation, multiple speakers, etc. Accordingly,  stronger techniques are needed to make the model generalize well to these adverse conditions.

Table~\ref{Table: prod long form} reports the long form WER when evaluating at 8-bit and 4-bit precision, for models trained without QAT, with variational noise (VN) using a constant Gaussian noise variance for all weights, 4-bit STE-QAT, and 4-bit PQN-QAT. Generally, 8-bit inference does not require QAT to perform on par with float models, and serves as an barometer for generalization behavior in this section. As expected, VN helps generalization and improves 8-bit WER from 17.2 to 16.2. Interestingly, VN, despite not being tuned to the unique quantization noise statistics in every channel in every tensor, also improves 4-bit WER from 27.1 to 20.6. However, we see that the 4-bit PQN-QAT shows the best WER at 8-bit and 4-bit, which implies that RAND with $L_\infty$ or some other $L_p$ norm can be useful for generaliztion as well. We leave that to future investigations.

\begin{table}
  \caption{Cascaded Conformers experiments for 4-bit QAT, evaluated at 4-bit and 8-bit, on a large scale in-house long form dataset.}
  \label{Table: prod long form}
  \centering
  \begin{tabular}{lllll}
    \toprule
    Eval precisions & Outlier method    & QAT method   & Long form WER \\
    \midrule
    Float / Int4 & None & None & 17.2 / 27.1   \\
    Int8 / Int4  & None & Tuned VN & 16.2 / 20.6   \\
     Int8 / Int4 & Norm & 4-bit STE & 16.1 / 17.0 \\
     Int8 / Int4 & Norm & 4-bit PQN & 15.4 / 15.6 \\
    \bottomrule
  \end{tabular}
\end{table}

\textbf{Multi-domain:}
Table~\ref{Table: prod} reports the 8-bit and 4-bit WER on short form and medium form utterances. Overall, the proposed norm decay with PQN-QAT (RAND) has 8-bit and 4-bit WER that are the closest to the float WER, and represents the best choice for these large-scale production datasets.

\begin{table}
  \caption{Cascaded Conformers experiments for 4-bit QAT, evaluated at 4-bit and 8-bit, on large scale in-house short form and medium form datasets.}
  \label{Table: prod}
  \centering
  \begin{tabular}{llllll}
    \toprule
    Eval precisions & Outlier method    & QAT method  & Short form WER & Medium form WER \\
    \midrule
    Float / Int4  & None & None & 4.9 / 7.4 & 3.7 / 8.7 \\
    Int8 / Int4  & None & Tuned VN & 5.0 / 5.7 & 3.8 / 5.5  \\
     Int8 / Int4 & Norm & 4-bit STE & 5.0 / 5.1 & 3.7 / 3.8 \\
     Int8 / Int4 & Norm & 4-bit PQN & 4.8 / 4.9 & 3.7 / 3.9 \\
    \bottomrule
  \end{tabular}
\end{table}

\textbf{Multi-precision:}
For vanilla STE-QAT without any outlier regularization, it tunes the network to perform the best at the precision that it is trained under. Even though intuitively 8-bit quantization is easier than 4-bit, Table~\ref{Table: vanilla ste} shows that the vanilla STE-QAT model has worse WER when running as 8-bit than as 4-bit. This implies that using the 4-bit aware model checkpoint at 8-bit or mixed precision (different layers at different precision) will have a big gap compared to training the model specifically for the inference time layer precision.

\begin{table}
  \caption{Cascaded Conformers experiments for 4-bit STE QAT without any outlier control method, showing the degraded 8-bit WER when not using RAND.}
  \label{Table: vanilla ste}
  \centering
  \begin{tabular}{llllll}
    \toprule
    Eval precisions & Outlier method    & QAT method  & Short form WER & Medium form WER \\
    \midrule
    Int8 / Int4 & None & 4-bit STE & 5.2 / 5.0 & 4.0 / 4.0    \\
    \bottomrule
  \end{tabular}
\end{table}

Table~\ref{Table: prod} shows that the models trained using norm decay have lower WER when operating in 8-bit mode and very little gap to the float model performance. This also implies that, in a mixed precision setup, increasing the number of layers running in 8-bit mode is more likely to monotonically improve WER, which is a desirable property to enable training multi-precision capable neural networks in one shot.

\subsection{Limitations}
While in this work we advocate for RAND with PQN, as with all noise based QAT, going to int2 or lower precision means that the increased variance of the training noise starts to hurt model convergence. Then, the design of the noise distribution becomes more critical. For example, for 2-bit symmetric quantization, the scale is the maximum weight entry. A large number of weights can change sign after introducing the noise with that high of a variance. Experiments show that this hurts network accuracy compared to if we prevented the sign switch from occurring. A principled way to stabilize PQN-QAT at 2-bit and beyond remains an open question.


Additionally, PQN-QAT in general does not follow the approach of what you train is what you serve~\cite{abdolrashidi2021pareto}: during training we use~\eqref{Eq: per channel noise} at higher precision, but during inference we use~\eqref{Eq: per channel} for true quantized operations. As a result, it limits the application of PQN-QAT when we want to migrate to low-bit quantized training.

\section{Conclusion}
In this paper, we formalize the noise based QAT technique by writing it in terms of the weight tensor's $L_p$ norm. The explicit dependence on the $L_p$ norm allows us to control for its magnitude within the end-to-end optimization (RAND). Through large scale experiments on multi-domain seq2seq tasks, we show that this method's performance is further improved by having a quantization scale for every channel. With per-channel scale and the norm decay active, RAND outperforms STE based QAT and methods that advocate for dedicated learnable scale parameters on a range of datasets. These improvements open up the possibility of exploiting some of the secondary benefits of RAND: improved generalization and multi-precision performance.

\clearpage

\section{Appendix}

\subsection{Machine Translation Experiments}

To show that the proposed method is working on other seq2seq models and modalities, we evaluate it on a WMT 2014 English-to-French translation task. 

\textbf{Model and Optimizer:} We implement the \textit{Base} Transformer model with 65M parameters reported in~\cite{vaswani2017attention}. The \textit{Base} Transformer model composed of encoder and decoder with N = 6 layers, model dimension $d_{model}$ = 512, hidden dimension $d_{ff}$ = 4$ \times d_{model}$ = 2048 and number of heads $h$ = 8.
We also evaluate its smaller version with model dimension $d_{model}$ = 256, hidden dimension $d_{ff}$ = 4$\times d_{model}$ = 1024 and number of heads $h$ = 4. This model has 30M parameters, we label it as \textit{Small}.

Above models use SentencePiece~\cite{sentencepiece} subword tokenizer with 32K vocabulary.

We use the same parameters for Adam optimizer as in~\cite{vaswani2017attention} but with \textit{warmup\_steps} = 4000. As in~\cite{vaswani2017attention} we use residual dropout with value 0.1, but with label smoothing of value 0.

\textbf{Training Data and Evaluation:} The models are trained on the WMT 2014 English-French training data. Then we select the best checkpoint using the dev data set English-French newstest2013, and report BLEU on English-French newstest2014 data set. BLEU is computed according to~\cite{post2018} with parameters~\cite{sacrebleu}: \textit{smooth\_method}="exp"; \textit{smooth\_value}=0.0; \textit{force}=False; \textit{lowercase}=False; \textit{tokenize}=intl; \textit{use\_effective\_order}=False. 

\textbf{Hardware:} The MT model is trained on TPU v3-32~\cite{TPU} for 100,000 iterations with batch size 2048. It takes nine hours to finish model training.

\textbf{Experimental Results:} We quantize MT transformer model weights of encoder, decoder and embeddings with int4. Bias is not quantized because it is negligible in comparison to the model weights. We use symmetric per-channel quantization. In table~\ref{Table: mt} we show BLEU scores of \textit{Base} and \textit{Small} models quantized with the different approaches and the improvements from incorporating RAND.

\begin{table}
  \caption{BLEU scores of \textit{Base} and \textit{Small} models for 4-bit symmetric weight quantization with per channel scale, on English-to-French newstest2014 tests.}
  \label{Table: mt}
  \centering
  \begin{tabular}{lllll}
    \toprule
    Eval precision & Outlier method    & QAT method     & \textit{Base} model BLEU & \textit{Small} model BLEU \\
    \midrule
    Float & None & None      & 40.5 & 38.4    \\
     Int4 & None & 4-bit STE & 39.9 & 37.6    \\
     Int4 & None & 4-bit PQN & 39.7 & 37.5    \\
     Int4 & Norm & 4-bit STE & 40.1 & 38.2    \\
     Int4 & Norm & 4-bit PQN & 40.0 & 38.2    \\
    \bottomrule
  \end{tabular}
\end{table}

\subsection{Mixed Scale QAT Mode}
Although the main focus of this work is on QAT mode~\eqref{Eq: inf norm}, mixed scale QAT mode~\eqref{Eq: large norm} can offer additional benefits by providing gradient signals to regularize multiple outliers per training iteration. This prevents other weight entries from becoming outliers as soon as one outlier is reguarlized by RAND. Mixed scale QAT mode show increased benefits for models that are either trained with per tensor scale, or very large models where each output channel has a large number of parameters.

\textbf{Experimental Results:} For experimentation, we set $p=8$ and $c = \frac{1}{2^{\text{bit}-1}-1}$. For training stability, we introduce another hyperparameter $k$, which controls the number of the weight entries with the largest magnitude that contribute to the norm calculation. This changes the mixed scale QAT mode equation into
\begin{equation}
\label{Eq: matmul large norm}
    Y_i = \left(W_i+ \frac{ \text{top\_k}(\norm{W_{i,j}}_p)}{2^{\text{bit}-1}-1} Z_i \right) X, 1 \leq i \leq M.
\end{equation}
All other settings are identical to the corresponding experiments from Section 5.

Table~\ref{Table: mixed scale} shows the comparison between per tensor scale with RAND, per tensor scaled with mixed scale RAND, and per channel scale with RAND. The LibriSpeech ConformerS WER gap between per tensor and per channel is closed by roughly half after introducing the mixed scale QAT mode.
\begin{table}
  \caption{ConformerS experiments for 4-bit symmetric weight quantization on LibriSpeech. The mixed scale is computed via the $8$-norm of the eight weight entries with the largest magnitude.}
  \label{Table: mixed scale}
  \centering
  \begin{tabular}{lllllll}
    \toprule
    Scale & Outlier method    & $k$ & QAT method  & Test-clean & Test-other \\
    \midrule
    Per tensor & Norm & N/A & 4-bit PQN & 2.71 & 6.50    \\
    Per tensor & Mixed & 8 & 4-bit PQN & 2.70 & 6.41    \\
    Per channel & Norm & N/A & 4-bit PQN & 2.64 & 6.30    \\
    \bottomrule
  \end{tabular}
\end{table}

Table~\ref{Table: mixed stew} shows the comparison between per channel scale with RAND and mixed scale RAND. The ConformerL's WER averaged across all datasets of SpeechStew is further reduced by introducing the mixed scale formulation in~\eqref{Eq: matmul large norm}.
\begin{table}
  \caption{Multi-domain ConformerL experiments for 4-bit symmetric weight quantization on SpeechStew dataset. CV: Common Voice; SB: Switchboard; LS: LibriSpeech; Ted: TED-LIUM; WSJ: Wall Street Journal. The mixed scale is computed via the $8$-norm of the two weight entries with the largest magnitude.}
  \label{Table: mixed stew}
  \centering
  \begin{threeparttable}
  \begin{tabular}{lll|cccccccc}
    \toprule
     Outlier    & QAT & \multirow{2}{*}{$k$} & \multicolumn{2}{c}{AMI} & \multirow{2}{*}{CV}  & \multicolumn{2}{c}{LS-Test} & \multirow{2}{*}{SB} & \multirow{2}{*}{Ted} & \multirow{2}{*}{WSJ}         \\
    \cmidrule(r){4-5} \cmidrule(r){7-8}
      Method   &   Method   & & IHM & SDM1 &  & clean & other \\
    \midrule
Norm & 4-bit PQN & N/A & 9.39 & 24.23 & 10.17 & 2.07 & 4.48 & 8.58 & 4.11 & 1.44 \\
Mixed & 4-bit PQN & 2 & 9.26 & 23.97 & 10.17 & 2.08 & 4.48 & 8.55 & 4.16 & 1.37 \\
    \bottomrule
  \end{tabular}
  \end{threeparttable}
\end{table}

\clearpage

\bibliographystyle{unsrt}
\bibliography{ref}

\end{document}